\documentstyle[aps,twocolumn,graphicx]{revtex}
\def\ave#1{\langle #1\rangle}
\newcommand{\ket}[1]{|#1\rangle}
\newcommand{\op}[1]{\hat{#1}}
\newcommand{\tr}{{\,\rm tr\,}}
\begin{document}

\title{Can quantum chaos enhance stability of quantum computation?}

\author{Toma\v z Prosen and Marko \v Znidari\v c}
\address{Physics Department, Faculty of Mathematics and Physics, 
University of Ljubljana, Slovenia}
\date{\today}
\draft
\maketitle

\begin{abstract}
We consider stability of a general quantum algorithm with respect to a fixed
but unknown residual interaction between qubits, and show a surprising fact, namely 
that the average fidelity of quantum computation increases by decreasing average time 
correlation function of the perturbing operator in sequences of consecutive quantum gates.  
Our thinking is applied to the quantum Fourier transformation where an alternative
'less regular' quantum algorithm is devised which is qualitatively more robust against
static random residual $n$-qubit interaction.
\end{abstract}


Recent investigations of theoretical and experimental possibilities of quantum information 
processing have made the idea of quantum computation\cite{feynman} very attractive and 
important (see e.g. \cite{review} for a review). Having the apparatus which is capable
of manipulation and measurement on pure states of individual quantum systems one can
make use of massive intrinsic parallelism of coherent quantum time evolution.

The main idea of quantum computation is the following: 
Consider a many-body system of $n$ elementary 
two-level quantum excitations --- {\em qubits}, which is called 
the {\em quantum register}, store the data for quantum computation in the initial 
state of a register $\ket{{\rm r}}$ which is a superposition of an exponential number 
${\cal N}=2^n$ of basic qubit states, then perform certain unitary transformation 
$U$ by decomposing $U=U(T)\cdots U(2)U(1)$ into a sequence of $T$ elementary 
one-qubit and two-qubit {\em quantum gates} $U(t)$, $t=1,2,\ldots,T$, such decomposition being
called a {\em quantum algorithm} (QA), and in the end obtain the
results by performing measurements of qubits on a final register state 
$U\ket{{\rm r}}$. QA is called {\em efficient} if the number of 
needed elementary gates $T$ grows with at most {\em polynomial} rate in $n = \log_2 {\cal N}$, 
and only in this case it can generally be expected to outperform the best
classical algorithms (in the limit $n\to\infty$). At present only few efficient QAs 
are known, and perhaps the most generally useful is the Quantum Fourier 
transformation (QFT) \cite{qft}. 

There are two major obstacles for performing practical quantum computation: First, there is a  
problem of {\em decoherence}\cite{decoherence} resulting from an unavoidable {\em time-dependent} coupling between qubits and the environment. If the perturbation couples only a small number of qubits at a time then such
errors can be eliminated at the expense of extra qubits by 
{\em quantum error correcting codes} \cite{error} (see Ref. \cite{TL} for another approach). Second, even if one knows an efficient error correcting code 
or assumes that quantum computer is ideally decoupled from the environment, 
there will typically exist a small {\em unknown} or {\em uncontrollable} 
residual interaction among qubits which one may describe by a general {\em static} perturbation. Therefore, understanding the {\em stability} of QAs with respect to various types 
of perturbations is an important problem (see \cite{banacloche,dimaKR,Zurek} for some results on 
this topic).

Motivated by \cite{peres}, we propose a new approach to the stability of quantum 
computation with respect to a static but incurable (perhaps unknown) perturbing interaction. 
We consider QA as a time-dependent dynamical system and relate 
its fidelity measuring the Hilbert space distance between computed
states of exact and perturbed algorithm in terms of integrated time-autocorrelation of the 
perturbing operator (generalizing Ref. \cite{kif}). 
The derived relation looks very surprising: it tells that faster decay of time-correlations of 
the perturbation between sequences of successive quantum gates means larger 
fidelity, and vice versa. We propose to use our rule of thumb as a guide to devise or to improve
QAs, either by introducing extra 'chaotic' gates or by
rewriting the gates in a different order in order to make time-evolution $U(t)$ 
'more chaotic'. As an important example, the well known QFT algorithm whose 
internal dynamics appears unpleasantly 'regular' has been improved 
in a way that the modified algorithm becomes qualitatively more robust against 
static random perturbation of the gates. We think our effect should be considered in experimental realization of QFT which are underway \cite{ExpOnQFT}. 
 
Let us write the {\em partial evolution operator} for a sequence of consecutive gates from 
$t'$ to $t$, $t' < t$, as $U(t,t') = U(t) U(t-1)\cdots U(t'+2) U(t'+1)$, with $U(t,t)\equiv 1$, 
and perturb the quantum gates by a (generally time-dependent) perturbation of strength $\delta$
generated by hermitean operators $V(t)$
\begin{equation}
U_\delta(t) = U(t)\exp(-i\delta V(t)).
\label{eq:pert}
\end{equation}
Propagating the initial register
state $\ket{{\rm r}}$ with exact and perturbed algorithms we focus on the {\em fidelity} of the
QA defined as
\begin{equation}
F(T) = \frac{1}{\cal N}\tr U_\delta^\dagger(T,0) U(T,0)
\label{eq:deffid}
\end{equation}
as an average over all initial register states.
Defining the Heisenberg time evolution from $t'$ to $t$, 
$V(t,t') = U^\dagger(t,t')V(t)U(t,t')$, we rewrite the fidelity as
\begin{equation}
F(T) = \frac{1}{\cal N}\tr\left( e^{i\delta V(1,0)}e^{i\delta V(2,0)} 
\cdots e^{i\delta V(T,0)}\right)
\label{eq:prodfid}
\end{equation}
by $T$ insertions of the unity $U^\dagger(t,0)U(t,0) = 1$ and observing 
$U^\dagger(t-1,0)U^\dagger_\delta(t)U(t,0) = \exp(i\delta V(t-1,0))$.

Next we make a series expansion in $\delta$ expressing the fidelity in terms of 
correlation functions
\begin{equation}
F(T) = 1 + \frac{1}{\cal N}\sum_{m=1}^\infty \frac{i^m \delta^m}{m!}
\!\!\!\!\!\!\sum_{t_1,\ldots,t_m=1}^{T}\!\!\! 
\tr\left(\op{\cal T}\prod_{j=1}^m V(t_j,0)\right).
\label{eq:serfid}
\end{equation}
where $\op{\cal T}$ is a left-to-right time ordering (w.r.t. indices $t_j$).
We can make the series starting at second order $m=2$ by assuming the average perturbation to be {\em traceless}, $\tr\bar{V}=(1/T)\sum_{t=1}^T \tr V(t,0) \equiv 0$ 
(otherwise, the effect of subtracting the trace average is a simple complex rotation of fidelity). To second order in $\delta$, the fidelity can be written as
\begin{equation}
F(T)=1-\frac{\delta^2}{2} \sum_{t,t'=1}^T  C(t,t') + {\cal O}(\delta^3),
\label{eq:Fcor}
\end{equation}
in terms of a 2-point time correlation (correlator) of the perturbation 
$C(t,t') := C(t',t) := \tr(V(t',0)V(t,0))/{\cal N} = \tr(V(t')V(t,t'))/{\cal N}$.
The relation (\ref{eq:Fcor}) is very interesting: it tells that the QA is
more stable if the time correlations of the perturbation are smaller on average, meaning that the
`chaotic' quantum time evolution is more stable than the `regular' one\cite{kif}.
One may use this general philosophy as a guide to design QAs, or to improve
the existing ones by rearranging quantum gates.

However, the behavior of the correlation function depends also on 
explicit time-(in)dependence of the perturbation $V(t)$. For example,
if the perturbation $V(t)$ is an {\em uncorrelated noise}, as would be in the case of 
coupling to an {\em ideal heath bath}, then the matrix elements of $V(t)$ may be assumed to be 
{\em gaussian random} with variances
$\ave{V_{jk}(t) V_{lm}(t')}_{\rm noise} = (1/{\cal N})\delta_{jm}\delta_{kl}\delta_{tt'}$. 
Hence one finds $\ave{C(t,t')}_{\rm noise} = \delta_{t t'}$, and averaging of a formula 
(\ref{eq:serfid}) yields the noise-averaged fidelity 
$\ave{F(T)}_{\rm noise} = \exp(-\delta^2 T/2)$ 
which is {\em independent of the QA} $U(t)$. On the other hand, 
for a {\em static} residual interaction $V(t)\equiv V$ one may 
expect slower correlation decay, depending on the 'regularity' of the 
evolution operator $U$, and hence faster decay of fidelity. 
Importantly, note that in a physical situation, where perturbation is expected to 
be a combination $V(t) = V_{\rm static} + V_{\rm noise}(t)$, the 
fidelity drop due to a static component
is expected to {\em dominate} long-time quantum computation $T\to\infty$ over the
noise component (e.g. due to decoherence), as soon as QA exhibits 
long time correlations of the operator $V_{\rm static}$.
Since the sequence of gates to accomplish a certain task $U$ is by no means unique, 
the natural question arises, how to write the QA in order to have fastest decay
of time correlations with respect to a static, say gaussian random (GUE) perturbation?
\par
We consider QFT working in a 
Hilbert space of dimension ${\cal N}=2^n$ with basis qubit states denoted by $\ket{k}, \ \ k=0,\ldots,2^n-1$. The 
unitary matrix $U_{\rm QFT}$ performs the following transformation on a state with 
expansion coefficients $x_k$
\begin{equation}
U_{\rm QFT} (\sum_{k=1}^{\cal N}{x_k \ket{k}})=\sum_{k=1}^{\cal N}{\tilde{x}_k\ket{k}},
\end{equation}
where $\tilde{x}_k=\frac{1}{\sqrt{{\cal N}}}\sum_{j=1}^{\cal N}{\exp{(2\pi i j k/{\cal N})} x_j}$. 
``Dynamics'' of QFT consists of three kinds of unitary gates: 
One-qubit gates ${\rm A}_j$ acting on $j$-th 
qubit
\begin{equation}
{\rm A}_j=\frac{1}{\sqrt{2}} \pmatrix{ 1 & 1 \cr
  1 & -1 \cr},
\label{eq:Agate}
\end{equation}
diagonal two-qubit gates ${\rm B}_{jk}={\rm diag}\{ 1,1,1,\exp{(i \theta_{jk})} \}$, 
with $\theta_{jk}=\pi/2^{k-j}$, and transposition gates ${\rm T}_{jk}$ which interchange 
$j$-th and $k$-th qubits. There are $n$ ${\rm A}$-gates, $n(n-1)/2$ ${\rm B}$-gates 
and $[n/2]$ transposition gates, where $[x]$ is an integer part of $x$. 
The total number of gates for the whole algorithm is therefore $T=[n(n+2)/2]$. 
For instance, in the case of $n=4$ we have a sequence of $T=12$ gates (time runs 
from right to left)
\begin{equation}
U_{\rm QFT} = 
{\rm T}_{03} {\rm T}_{12} {\rm A}_0 {\rm B}_{01} {\rm B}_{02} {\rm B}_{03} {\rm A}_1 {\rm B}_{12} {\rm B}_{13} {\rm A}_2 {\rm B}_{23} {\rm A}_3.
\label{eq:plain4}
\end{equation}  
\par
In what follows we will focus on a static random perturbation, that 
is $V(t) \equiv V$ is a random ${\cal N}$-dimensional GUE matrix 
with normalized second moments $\ave{V_{jk}V_{lm}}=\delta_{jm}\delta_{kl}/{\cal N}$, where 
$\ave{.}$ denotes an average over GUE. For small perturbation strength $\delta$ the quantity 
controlling the fidelity (\ref{eq:Fcor}) is the correlator
\begin{equation}
\ave{C(t,t')} = \frac{1}{\cal N}\ave{\tr(V(t,t')V)} = 
\left|\frac{1}{{\cal N}} \tr{U(t,t')}\right|^2.
\label{eq:C_anal}
\end{equation}
Averaging over GUE is done only to ease up analytical calculation and to yield a 
quantity that is independent of a particular realization of perturbation. 
Qualitatively similar (numerical) results are obtained without the averaging. 
We have $\ave{C(t,t)}\equiv 1$ due to normalization of the 
second moments of GUE, while for an arbitrary fixed $V$, the diagonal correlator is
\begin{equation}
C(t,t)=C(0,0)=\frac{1}{{\cal N}} \tr{V^2}.
\label{eq:C00}
\end{equation}
In a sum of correlation function (\ref{eq:Fcor}) we must therefore distinguish two 
contributions: (i) The diagonal correlator (\ref{eq:C00}) 
just sets an overall scale. This is a {\em static} quantity as it depends on the strength 
of a perturbation $V$ only and can be included in $\delta$ by normalizing $\tr{V^2}/{\cal N}=1$. 
(ii) The off-diagonal contribution is mainly determined by the rate of decay of $C(t,t')$ as $t-t'$ 
increases which is an essential {\em dynamical} feature of QA.
\par
\begin{figure}
\begin{center}
\begin{minipage}[b]{2.1in}
\includegraphics[height=2.1in]{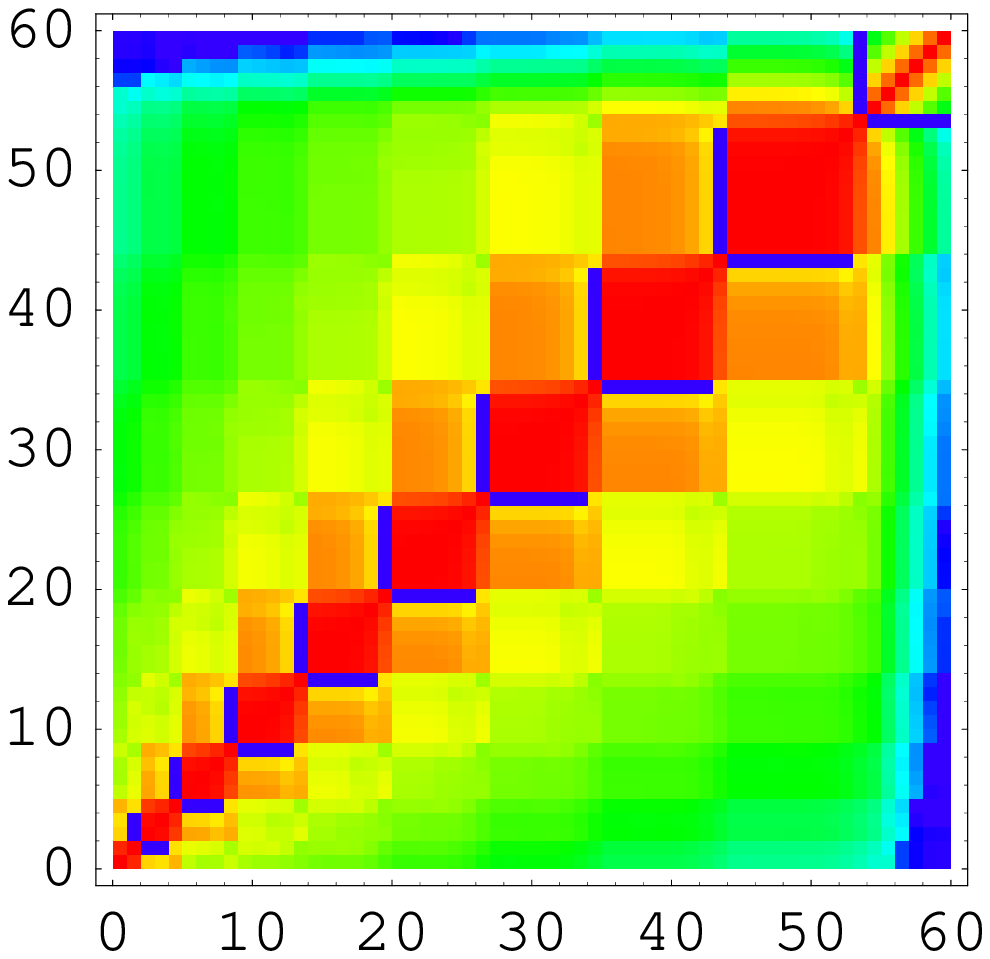}
\end{minipage}
\hspace{0mm}
\begin{minipage}[b]{0.4in}
\includegraphics[height=2.1in]{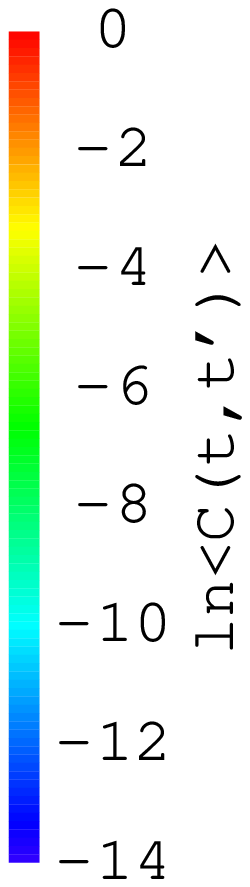} 
\end{minipage}
\end{center}
\vspace{-2mm}
\centerline{\includegraphics[width=2.57in]{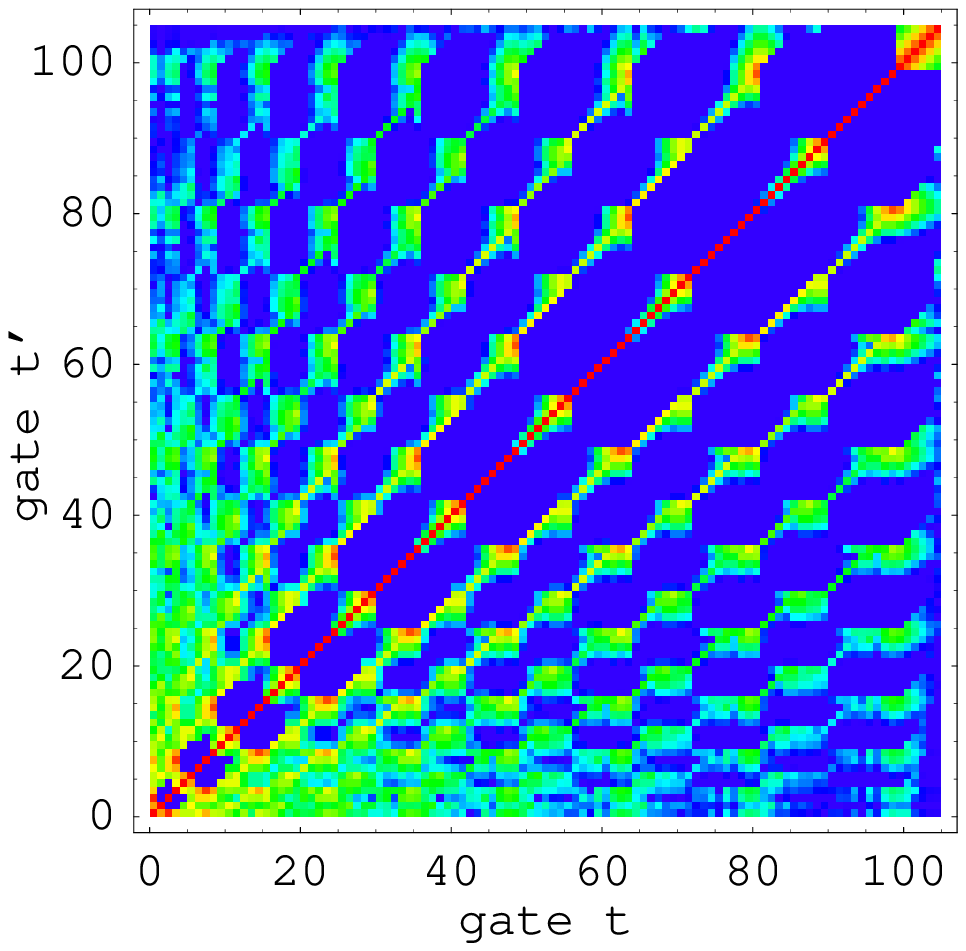} \hspace{7mm}}
\caption{Correlation function $\ave{C(t,t')}$ for $n=10$ qubits and GUE perturbation. 
Top figure shows standard QFT (\ref{eq:plain4}) with $T=60$, 
while bottom figure shows IQFT with $T=105$ gates. Color represents the size of 
elements in log-scale from red ($e^{-0}$) to blue ($e^{-14}$ and less).}
\label{fig:corel}
\end{figure}

We have calculated the correlator $\ave{C(t,t')}$ for QFT (\ref{eq:C_anal}) 
which is shown in top fig.\ref{fig:corel}. One can clearly see square red plateaus 
on the diagonal due to blocks of successive ${\rm B}$-gates. 
Similar square plateaus can also be seen off diagonal (from orange, yellow to green), so that 
the correlation function has a staircase-like structure, with the ${\rm A}$-gates responsible for 
the drops and ${\rm B}$-gates responsible for the flat regions in between. 
This can be easily understood. For ``distant'' qubits $k-j \gg 1$ 
the gates ${\rm B}_{jk}$ are close to the identity and therefore cannot reduce the correlator. This slow correlation decay results in the correlation sum 
$\chi:=\frac{1}{2} \sum_{t,t'=1}^T C(t,t')$ 
being proportional to $\chi \propto n^3$ (sum of the first $n$ squares) as 
compared to the theoretical minimum $\chi \propto T \propto n^2$. 
\par
\begin{figure}
\centerline{\includegraphics[width=3.3in]{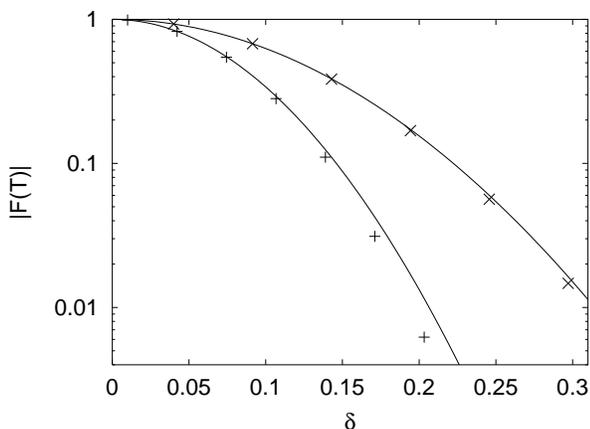}}
\caption{Dependence of fidelity $|\ave{F(T)}|$ on $\delta$ for QFT (pluses) and IQFT (crosses), for fixed
$n=8$. 
Solid curves are functions $\exp{(-\chi \delta^2)}$ (see text) with $\chi$ calculated analytically 
(\ref{eq:C_anal}) and equal to 
$\chi=108$ for QFT and $46.6$ for IQFT.}
\label{fig:gauss}
\end{figure}

In view of this, we will now try to rewrite the QFT with a goal to accomplish 
$\chi \propto n^2$. From (\ref{eq:C_anal}) we learn that the gates that are {\em traceless} 
(e.g. ${\rm A}$-gates) reduce the correlator very effectively. In the plain QFT algorithm 
(\ref{eq:plain4}) we have $n-1$ blocks of ${\rm B}$-gates, where in each block all ${\rm B}$-gates 
act on the same first qubit, say $j$. In each such block, we propose to replace ${\rm B}_{jk}$ 
with a new gate ${\rm G}_{jk}={\rm R}_{jk}^\dagger {\rm B}_{jk}$, 
where unitary gate ${\rm R}_{jk}$ will be chosen so as to 
commute with all the diagonal gates ${\rm B}_{jl}$ in the block, whereas at the end of the 
block we will insert ${\rm R}_{jk}$ in order to ``annihilate'' ${\rm R}^\dagger_{jk}$ 
so as to preserve the evolution matrix of a whole block. 
Unitarity condition ${\rm R}^\dagger_{jk} {\rm R}_{jk}=1$ and 
$[{\rm R}_{jk},{\rm B}_{jl}]=0$ for all $j,k,l$ leave us with a 6 parametric set 
of matrices ${\rm R}_{jk}$. 
By further enforcing $\tr{{\rm R}_{jk}}=0$ in order to maximally reduce the correlator, 
we end up with 4 free real parameters in ${\rm R}_{jk}$. One of the simplest choices, that has  
been proved to be equally suitable as any other, is the following
\begin{equation}
{\rm R}_{jk}=\pmatrix{ 0 & 0 & -1 & 0 \cr
		0 & 1 & 0 & 0 \cr
		1 & 0 & 0 & 0 \cr
		0 & 0 & 0 & -1 \cr}.
\label{eq:R}
\end{equation}
Furthermore, we find that ${\rm R}$-gates also commute among themselves, 
$[{\rm R}_{jk},{\rm R}_{jl}]=0$, which enables us to write a sequence of
${\rm R}$-gates as we like, e.g. in the same order as a sequence of ${\rm G}$'s, so 
that pairs of gates ${\rm G}_{jk}$, ${\rm R}_{jk}$ operating on same pair of qubits 
$(j,k)$, whose product is a {\em bad gate} ${\rm B}_{jk}$, are never neighboring.
This is best illustrated by an example. For instance, the block 
${\rm B}_{01} {\rm B}_{02} {\rm B}_{03}$ will be replaced by 
${\rm R}_{01} {\rm R}_{02} {\rm R}_{03}
{\rm R}^\dagger_{01}{\rm B}_{01} {\rm R}_{02}^\dagger{\rm B}_{02} {\rm R}^\dagger_{03} {\rm B}_{03} 
= {\rm R}_{01} {\rm R}_{02} {\rm R}_{03} {\rm G}_{01} {\rm G}_{02} {\rm G}_{03}$. 
This is how we construct an {\em improved Fourier transform algorithm} (IQFT). 
For IQFT we need one additional type of gates, instead of diagonal ${\rm B}$-gates, 
we use nondiagonal ${\rm R}$ and ${\rm G}$.
To illustrate the obvious general procedure we write out the whole IQFT algorithm for $n=4$ qubits 
(compare with (\ref{eq:plain4}))
\begin{eqnarray}
U_{\rm IQFT} = 
&&{\rm T}_{03} {\rm T}_{12} {\rm A}_0 {\rm R}_{01} {\rm R}_{02} {\rm R}_{03} {\rm G}_{01} {\rm G}_{02} {\rm G}_{03} {\rm A}_1 \nonumber \\
&&{\rm R}_{12} {\rm R}_{13} {\rm G}_{12} {\rm G}_{13} {\rm A}_2 {\rm R}_{23} {\rm G}_{23} {\rm A}_3.
\label{eq:IQFT4}
\end{eqnarray}
\begin{figure}
\centerline{\includegraphics[width=3.3in]{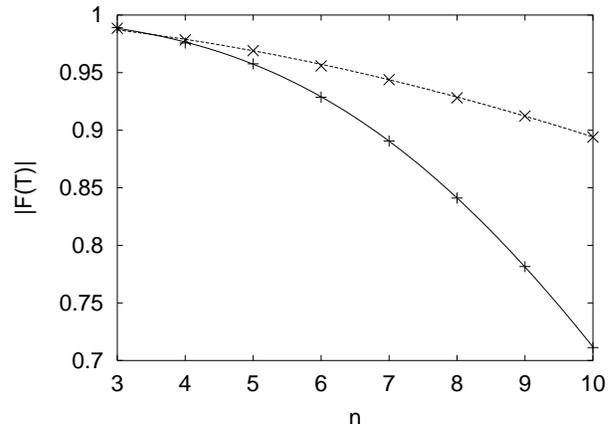}}
\caption{Dependence of fidelity $|\ave{F(T)}|$ on the number of 
qubits $n$ for QFT (pluses) and IQFT algorithms (crosses), for fixed $\delta=0.04$. 
Numerical averaging over 50 GUE realizations is performed. Full curve is 
$\exp{(-\delta^2 \{ 0.236 \ n^3 -0.38 n^2+1.45 n \})}$ and dashed is 
$\exp{(-\delta^2 \{ 0.61 n^2+0.89 n\})}$. For $n=10$ the trace is approximated by an average 
over 200 gaussian random register states.}
\label{fig:ft}
\end{figure}
Such IQFT algorithm consists of a total $T=[n(2n+1)/2]$ gates (note that it does not pay of 
to replace block with a single ${\rm B}$ gate as we have done, so we could safely leave 
${\rm B}_{23} \equiv {\rm R}_{23} {\rm G}_{23}$). 
The correlation function for IQFT algorithm is shown in bottom fig.\ref{fig:corel}. 
Almost all off-diagonal correlations are greatly reduced (to the level 
$\propto 1/{\cal N}^2$), leaving us only with a dominant diagonal. If we had 
only diagonal elements, the fidelity would be $\ave{F(T)}=1-\delta^2 \frac{T}{2}$,
(as in the case of noisy perturbation or decoherence, however, with a different physical meaning of the strength scale $\delta$) where the number of gates scales as $T \propto n^2$. From the pictures 
(\ref{fig:corel}) it is clear that we have very fast correlation decay for IQFT so that the
correlation sum $\chi$ has decreased from 
$\chi \propto n^3$ to $\chi \propto n^2$ behavior. To further illustrate this, 
we have numerically calculated the fidelity by simulating QA
and applying perturbation $\exp{(-i \delta V)}$ at each gate. The results are shown in 
fig.\ref{fig:ft}. Difference between $\propto n^2$ and $\propto n^3$ behavior is nicely seen. 
As we have argued before, the sum of 2-point correlator (\ref{eq:Fcor}) gives us only the 
first nontrivial order in $\delta$-expansion. 
For dynamical systems, being either integrable or mixing and ergodic, it has been 
shown\cite{kif} that also higher orders of (\ref{eq:serfid}) can typically be written as simple powers 
of the correlation sum $\chi$, so that the fidelity has a simple functional form 
$F(t)=\exp(-\chi\delta^2)$. Although QA has quite inhomogeneous time-dependence, we may still hope 
that $\exp{(-\chi \delta^2)}$ is a reasonable approximation to the fidelity also at
higher orders in $\delta$. This is in fact the case as can be seen in 
fig.\ref{fig:gauss}. 
Note also that the leading coefficient in the exponent for IQFT, 
$\lim_{n\to\infty}\chi/n^2 = 0.61$, is close to the theoretical minimum of $0.5$.
\par
As the definition of what is a fundamental single gate is somehow arbitrary, the problem of minimizing the sum $\chi$ depends on a given technical realization of gates and the nature of the residual perturbation $V$ 
for an experimental setup. We should mention that the optimization becomes harder if we consider 
{\em few-body} (e.g. two-body random) perturbation. This is connected with 
the fact that quantum gates are two-body operators and can perform only a very limited set 
of rotations on a full Hilbert space and consequently have a limited capability of reducing 
correlation functions in a single step. However, our simple approach based on $n$-body random matrices
seems reasonable if errors due to unwanted few-body qubit interactions can be eliminated by other 
methods\cite{error,TL}. 
\par
In conclusion, we have presented a novel approach to the stability of time-dependent quantum dynamics 
applied to the fidelity of quantum computation. For an uncorrelated time-dependent perturbation, the decay of 
fidelity does not depend on dynamics, however, for a static perturbation characterizing faulty gates the system is 
more stable, as reflected in a higher fidelity, the more ``chaotic'' it is and the faster 
correlations decay it has. Our idea is demonstrated on example of QFT algorithm perturbed by a
GUE matrix, devising an alternative QFT which is qualitatively more robust against
a static random perturbation of the gates. It is an interesting question how this dynamical enhancement of stability relates to ``chaotic melting'' of a static quantum computer \cite{casati}.

Discussions with T.~H.~Seligman are gratefully acknowledged. The work has been supported by the 
ministry of Education, Science and Sport of Slovenia.

\end{document}